\documentclass[twoside]{dis08}
\usepackage[latin1]{inputenc}
\usepackage[dvips]{graphicx,epsfig,color}
\usepackage{wrapfig,rotating}
\usepackage{amssymb,amsmath,array}

\newcommand{\be}{\begin{equation}}
\newcommand{\ee}{\end{equation}}
\newcommand{\bea}{\begin{eqnarray}}
\newcommand{\eea}{\end{eqnarray}}

\def\kt{k_\perp}

\def\kt{k_\perp}

\def\pp{p_\perp}

\def\avk{\langle k_\perp ^2\rangle}
\def\avp{\langle p_\perp ^2\rangle}
\def\avk{\langle k_\perp ^2\rangle}
\def\avp{\langle p_\perp ^2\rangle}

\def\T{_{_T}}
\def\C{_{_C}}

\def\lsim{\mathrel{\rlap{\lower4pt\hbox{\hskip1pt$\sim$}}\raise1pt\hbox{$<$}}}
\def\gsim{\mathrel{\rlap{\lower4pt\hbox{\hskip1pt$\sim$}}\raise1pt\hbox{$>$}}}
\def\nostrocostruttino#1\over#2{\mathrel{\mathop{\kern 0pt \rlap
{\hbox{$#1$}}} \hbox{\kern-.135em $#2$}}}

\begin{document}
\title{Transversity and Collins Fragmentation Functions: Towards a New Global 
Analysis}

\author{M.~Anselmino$^1$, M.~Boglione$^1$, U.~D'Alesio$^{2,3}$, 
A.~Kotzinian$^4$,\\ S.~Melis$^1$, F.~Murgia$^3$, {A.~Prokudin}$^{1,5}$
\footnote{Talk presented by A. Prokudin at the XVI International Workshop on 
Deep-Inelastic Scattering and Related Subjects, 
DIS 2008, London}, C.~T\"{u}rk$^1$
%
%
%
\vspace{.3cm}\\
%
1- Dipartimento di Fisica Teorica, Universit\`a di Torino and \\ 
          INFN, Sezione di Torino, Via P. Giuria 1, I-10125 Torino, Italy
\vspace{.1cm}\\
2- Dipartimento di Fisica, Universit\`a di Cagliari,\\ 
Cittadella Universitaria di Monserrato, I-09042 Monserrato (CA), Italy
\vspace{.1cm}\\
3- INFN, Sezione di Cagliari, C.P. 170, I-09042 Monserrato (CA), Italy
\vspace{.1cm}\\
4- CEA-Saclay, IRFU/Service de Physique Nucléaire, 91191 Gif-sur-Yvette, 
France; \\ Yerevan Physics Institute, 375036 Yerevan, Armenia;
JINR, 141980 Dubna, Russia
\vspace{.1cm}\\
5- Di.S.T.A., Universit\`a del Piemonte Orientale
             ``A. Avogadro'', Alessandria, Italy
}

\maketitle

\begin{abstract}
A new, preliminary global analysis of the experimental data on azimuthal 
asymmetries in SIDIS from HERMES and COMPASS collaborations, and in 
$e^+e^- \to h_1 h_2 X$
processes from the BELLE collaboration, is performed. 
The new data allow for a more precise determination of the Collins fragmentation 
function and of the transversity distribution function for
$u$ and $d$ quarks, in comparison with the results of our previous analysis.
Estimates for the single spin asymmetry
$A_{UT}^{\sin(\phi_h + \phi_S)}$ at JLab and
COMPASS, operating on a transversely polarized proton target, are presented.
\end{abstract}

The transversity distribution function, usually denoted as
$h_1$ or $\Delta_T q$, together with the unpolarized
distribution functions $q$ and the helicity distributions
$\Delta q$, contain basic information about the spin structure of 
nucleons.

SIDIS processes represent a testing ground for the transversity distribution 
where this chiral-odd function couples to another chiral-odd 
quantity, the Collins fragmentation function~\cite{Collins:1992kk}, giving 
rise to an azimuthal asymmetry, the so called Collins asymmetry.
Another azimuthal asymmetry produced in the $e^+e^- \to h_1 h_2 X$
process~\cite{Boer:1997mf} results from a convolution of two Collins 
fragmentation functions for quarks and antiquarks. 

The first global analysis of data on azimuthal asymmetries in SIDIS
and in $e^+e^-$ annihilations, resulting in the first extraction of the 
transversity distribution and the Collins fragmentation function, was 
presented in Ref.~\cite{Anselmino:2007fs}. 
Recently, much higher statistics data 
on the $A_{UT}^{\sin(\phi_{h}+\phi_S)}$ azimuthal asymmetries for SIDIS have 
become available: in Ref.~\cite{Diefenthaler:2007rj} the HERMES Collaboration 
have presented pion and kaon azimuthal asymmetries; 
in Ref.~\cite{Alekseev:2008dn} the COMPASS Collaboration have presented their 
measurements for separated charged pion and kaon asymmetries;
the BELLE collaboration have issued new high precision data of the  
Collins asymmetry in  $e^+e^-$ annihilation~\cite{Seidl:2008xc}. 
It is therefore timely to reconsider the results of 
Ref.~\cite{Anselmino:2007fs} using the new data in order to improve our 
knowledge of the Collins fragmentation function and of the transversity 
distribution.


\begin{table}[t]
\begin{center}
\begin{tabular}{|llll|}
\hline
$N_{u}^T$ =  $0.79 \pm 0.11$ & $N_{d}^T$ =  $ -1.00  \pm  0.15$ &
$\alpha$ =  $0.62  \pm  0.18$  & $\beta$  = $0.31  \pm  0.27$   \\
\hline 
$N_{fav}^C$  = $0.43  \pm  0.05$ & $N_{unf}^C$   = $-1.00  \pm  0.17$ &
$\gamma$  = $0.96  \pm  0.06$ & $\delta$   = $0.01  \pm  0.03$    \\
\hline 
$M^2_h$  =  $0.91 \pm 0.46 \;{\rm (GeV}/c)^2$ & & & \\
\hline
\end{tabular}
\end{center}
\vskip -0.5cm
\caption{Best values of the free parameters for the $u$ and $d$
transversity distribution functions and for the
favoured and unfavoured Collins fragmentation functions. We obtain 
$\chi^2/{\rm d.o.f.}\;\;$ = $1.3$.  Notice that
the errors generated by MINUIT are strongly correlated, and should not
be taken at face value. The significant fluctuations in our results
are shown by the shaded areas in Figs.~\ref{fig:hermes_compass} and 
\ref{fig:belle}.
\label{fitpar}}
\end{table}

We assume the usual parameterization of unpolarized distribution and 
fragmentation functions~\cite{Anselmino:2007fs}:
$
 f_{q/p}(x,\kt)=
   f_{q/p}(x)\;\frac{e^{-{\kt^2}/\avk}}{\pi\avk}\; ,\;\;\;
 D_{h/q}(z,\pp)=D_{h/q}(z)\;\frac{e^{-\pp^2/\avp}}{\pi\avp}\;,
$
with the values of $\langle k_\perp^2\rangle$ and $\langle p_\perp^2\rangle$ 
fixed to the values found in Ref.~\cite{Anselmino:2005nn} by analysing 
unpolarized SIDIS: 
$\langle\kt^2\rangle   = 0.25  \;({\rm GeV}/c)^2 \>,
\langle p_\perp^2\rangle  = 0.20 \;({\rm GeV}/c)^2 \>.$
Integrated parton
distribution and fragmentation functions $f_{q/p}(x)$ and $D_{h/q}(z)$ are 
available in the literature; in particular we will refer to 
Refs.~\cite{Gluck:1998xa} and~\cite{deFlorian:2007aj}.

As in  our previous analysis \cite{Anselmino:2007fs} we adopt the following 
parameterizations for the transversity distribution, $\Delta_T q(x, \kt)$, 
and the Collins FF, $\Delta^N D_{h/q^\uparrow}(z,\pp)$:
\bea
&& \Delta_T q(x, \kt) =
\frac{1}{2} \, {\cal N}^{\T}_q(x)\,
[f_{q/p}(x)+\Delta q(x)] \;
\frac{e^{-{\kt^2}/{\avk\T}}}{\pi \avk \T}\;, \label{tr-funct} \\
&& \Delta^N D_{h/q^\uparrow}(z,\pp) = 2\,{\cal N}^{\C}_q(z)\;
D_{h/q}(z)\;h(\pp)\,\frac{e^{-\pp^2/{\avp}}}{\pi \avp}\;,
\label{coll-funct}
\eea
with
\bea
&& {\cal N}^{\T}_q(x)= N^{\T}_q \,x^{\alpha} (1-x)^{\beta} \,
\frac{(\alpha + \beta)^{(\alpha +\beta)}} {\alpha^{\alpha} \beta^{\beta}}\,,
\;\;\;\;\;
{\cal N}^{\C}_q(z)= N^{\C}_q \, z^{\gamma} (1-z)^{\delta} \,
\frac{(\gamma + \delta)^{(\gamma +\delta)}}
{\gamma^{\gamma} \delta^{\delta}}\,,\\
&&
h(\pp)=\sqrt{2e}\,\frac{p_\perp}{M_{h}}\,e^{-{p_\perp^2}/{M_{h}^2}}\,,
\eea
and $-1\le N^{\T}_q\le 1$, $-1 \le N^{\C}_q \le 1$. We assume $\avk \T = \avk$.
The helicity distribution $\Delta q(x)$ is taken from Ref.~\cite{Gluck:2000dy}.

\begin{figure}[ht]
\vspace*{-.5cm}
\begin{center}
\includegraphics[width=0.4\textwidth,bb= 10 140 540 660,angle=-90]
{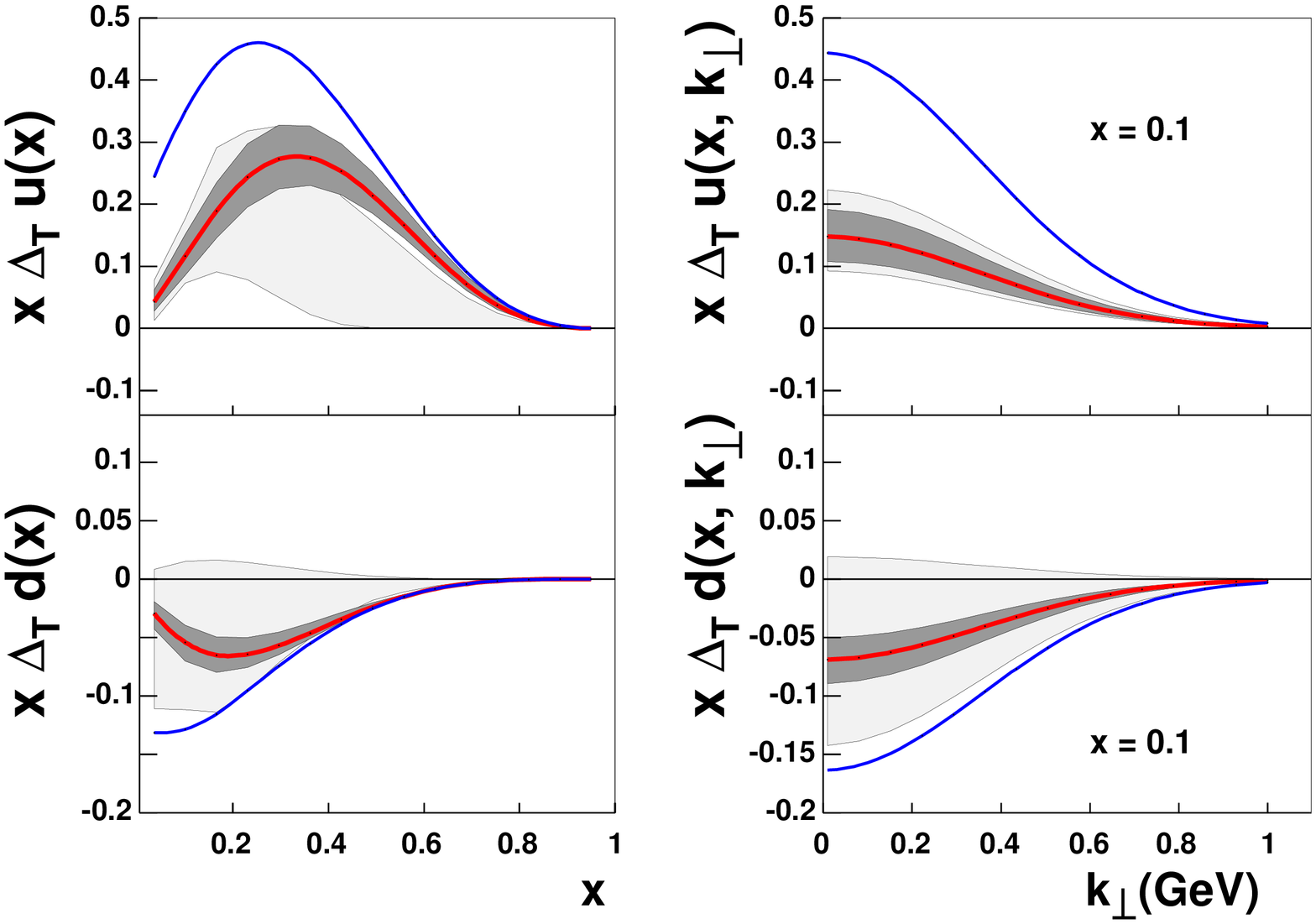}\hskip 2cm
\includegraphics[width=0.4\textwidth,bb= 10 140 540 660,angle=-90]
{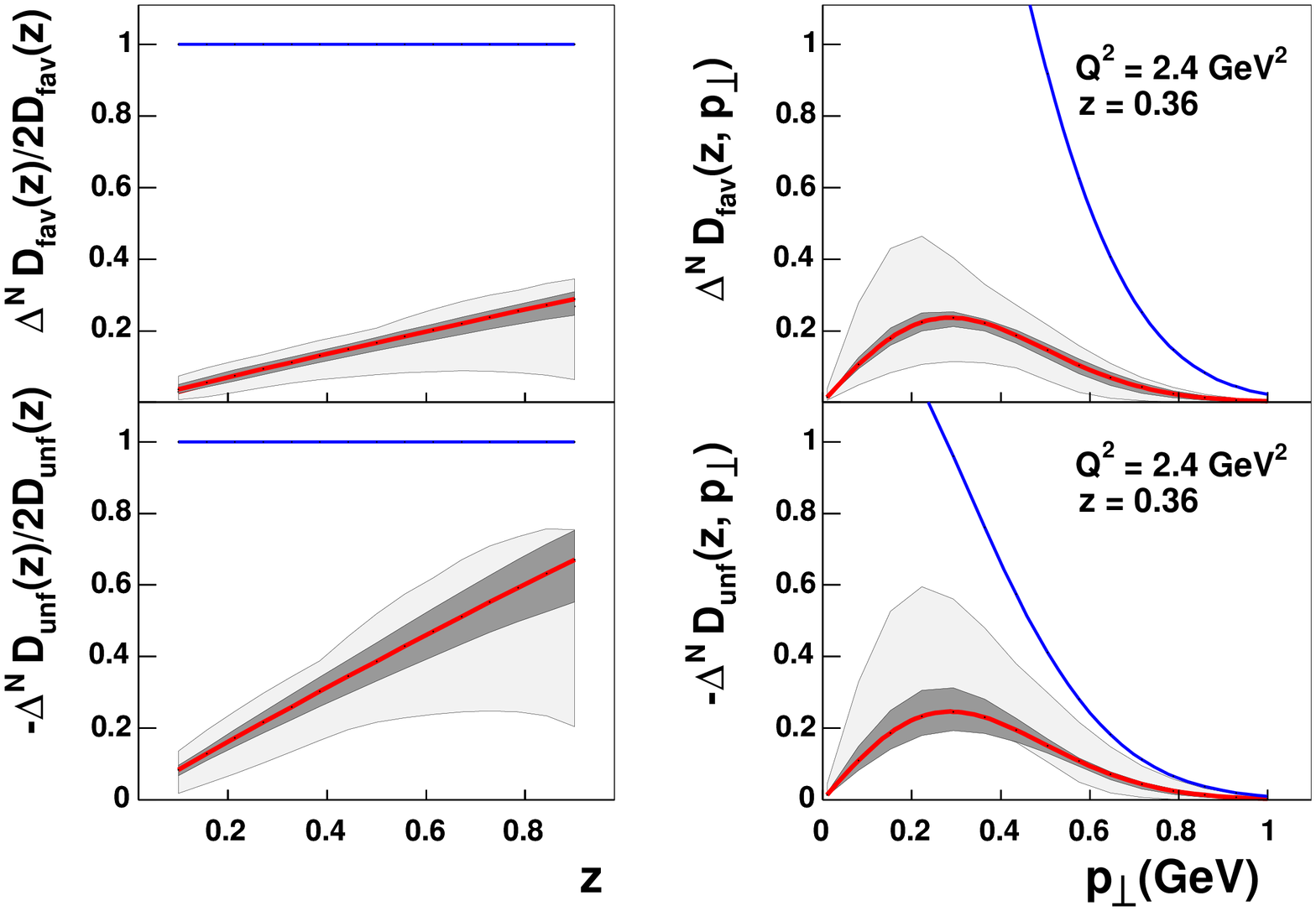}
\end{center}
\vskip -0.3cm
\caption{\label{fig:transv-coll}
Left panel: the transversity distribution functions for $u$ and $d$ flavours 
as determined by our global fit; we also show the Soffer bound 
(highest or lowest lines) 
and the (wider) bands of our previous extraction~\cite{Anselmino:2007fs}. 
Right panel: favoured and unfavoured Collins fragmentation functions as 
determined by our global fit; we also show the positivity bound and the 
(wider) bands as obtained in Ref.~\cite{Anselmino:2007fs}.}
\end{figure}

\begin{figure}[ht]
\vspace*{-1.7cm}
\begin{center}
\includegraphics[width=0.4\textwidth,bb= 10 140 540 660,angle=-90]
{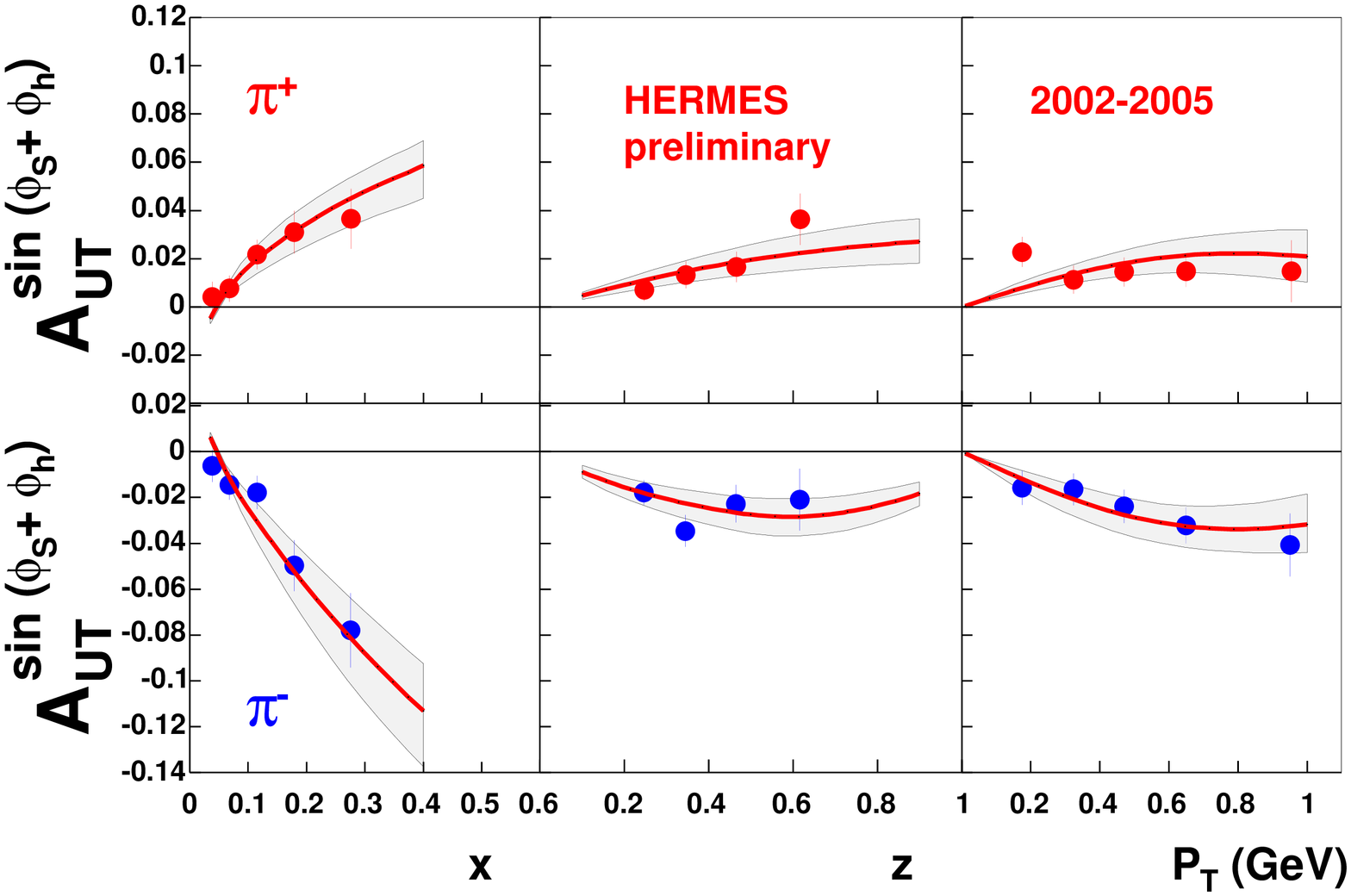} \hskip 2.5cm
\includegraphics[width=0.4\textwidth,bb= 10 140 540 660,angle=-90]
{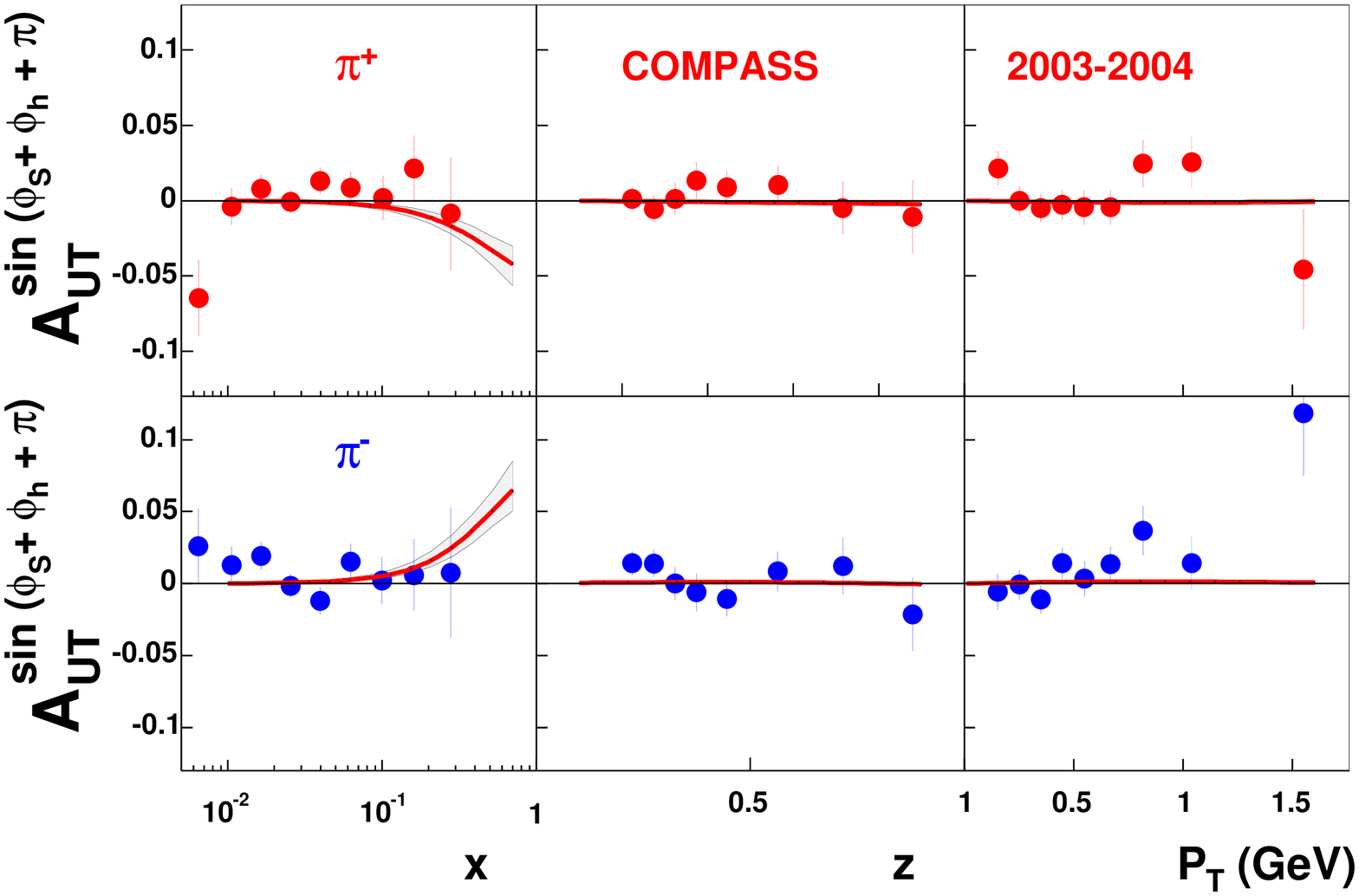}
\end{center}
\vskip -0.5cm
\caption{\label{fig:hermes_compass}
Fits of HERMES~\cite{Diefenthaler:2007rj} and COMPASS~\cite{Alekseev:2008dn} 
data. 
The shaded area corresponds to the 
uncertainty in the parameter values, see Ref.~\cite{Anselmino:2007fs}.}
\end{figure}

\begin{figure}[ht]
\vspace*{-1.7cm}
\begin{center}
\hspace*{1.5cm}
\includegraphics[width=0.42\textwidth,bb= 10 140 540 660,angle=-90]
{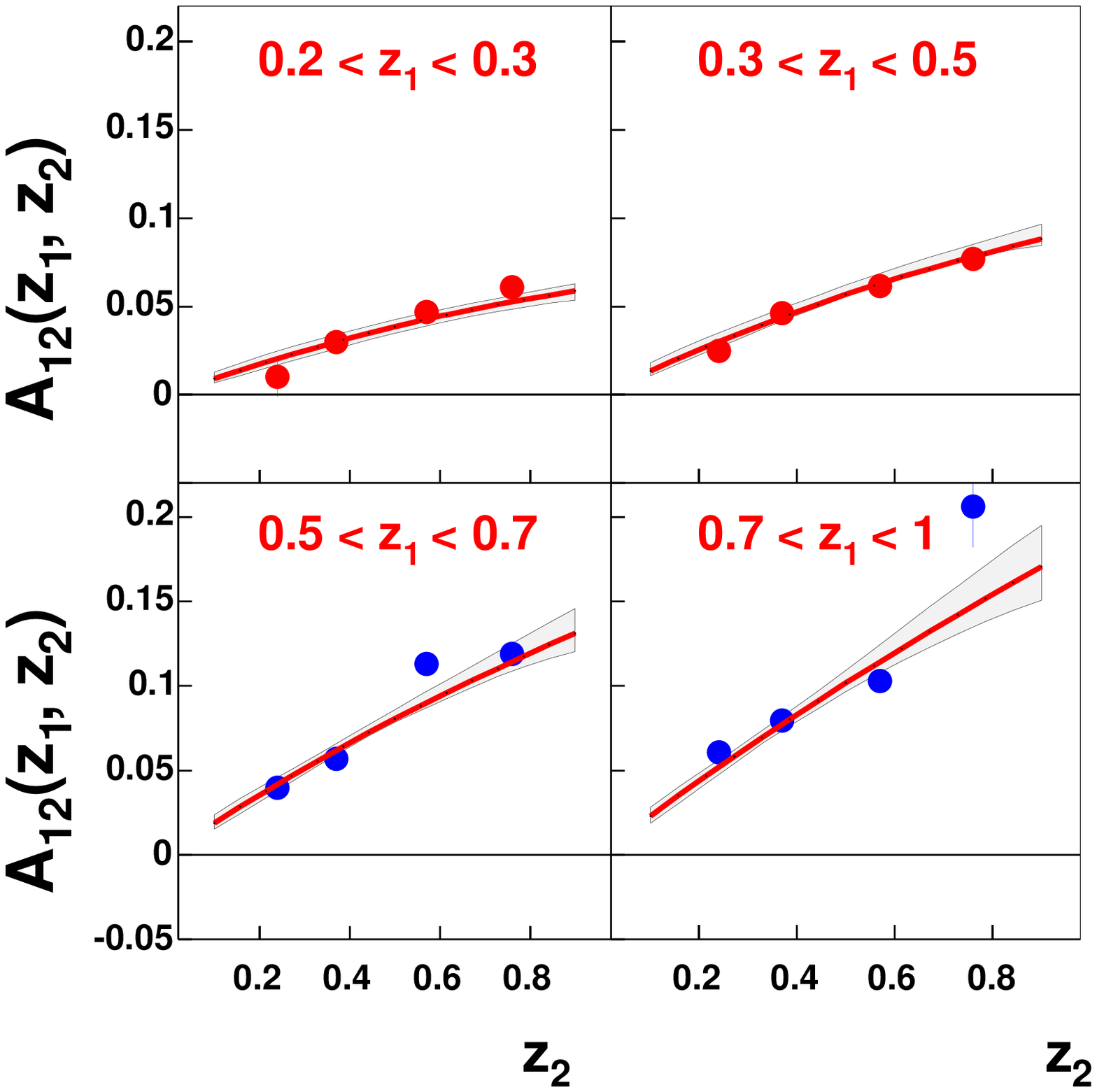} \hspace*{0.3cm}
\includegraphics[width=0.42\textwidth,bb= 10 140 540 660,angle=-90]
{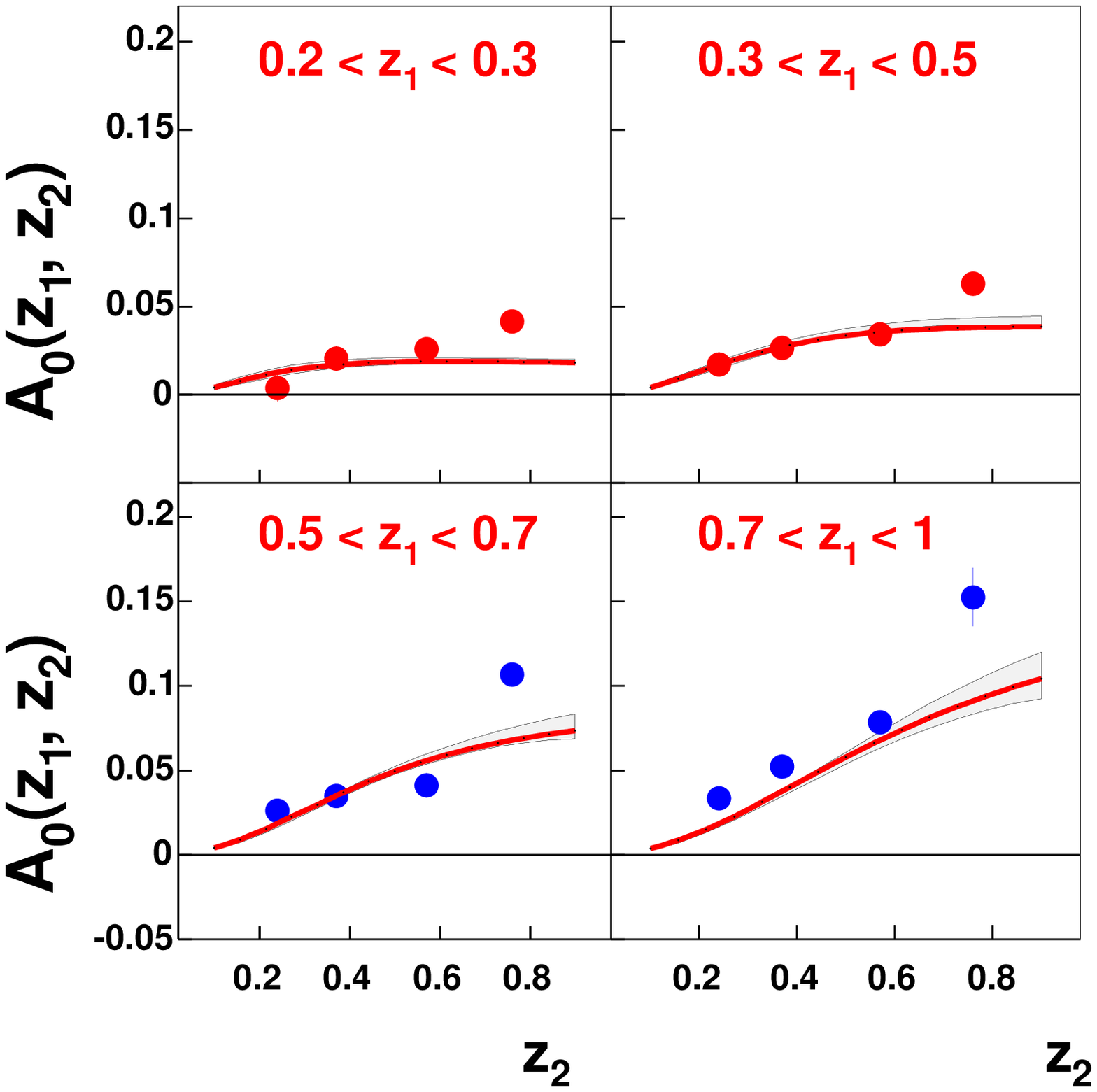}
\end{center}
\vskip -0.5cm
\caption{\label{fig:belle}
Left panel: fit of the BELLE~\cite{Seidl:2008xc} data on the $A_{12}$ 
asymmetry ($\cos (\varphi_1 + \varphi_2)$ method). 
Right panel: predictions for the  $A_{0}$ BELLE asymmetry 
($\cos(2\,\varphi_0)$ method).
}
\end{figure}

Table~\ref{fitpar} contains the results of the best fit to the new data sets, 
Refs.~\cite{Diefenthaler:2007rj,Alekseev:2008dn,Seidl:2008xc}. 
In Fig.~\ref{fig:transv-coll}, our present results for the transversity 
distribution and the Collins fragmentation function 
are compared to those of our previous extraction, 
Ref.~\cite{Anselmino:2007fs}; 
Fig.~\ref{fig:hermes_compass} shows the fits to the  
HERMES~\cite{Diefenthaler:2007rj} and 
COMPASS~\cite{Alekseev:2008dn} data, while Fig.~\ref{fig:belle} shows 
the fit to the BELLE $A_{12}$ asymmetry data and the predictions for the 
BELLE $A_{0}$ asymmetry \cite{Seidl:2008xc} (notice that we do not include the 
$A_{0}$ data in the fit, as the two sets of BELLE data are not indepedent;  
the consequences of fitting $A_{0}$ instead of $A_{12}$ are presently 
under investigation).

\begin{figure}[ht]
\vspace*{-1.7cm}
\begin{center}
\includegraphics[width=0.4\textwidth,bb= 10 140 540 660,angle=-90]
{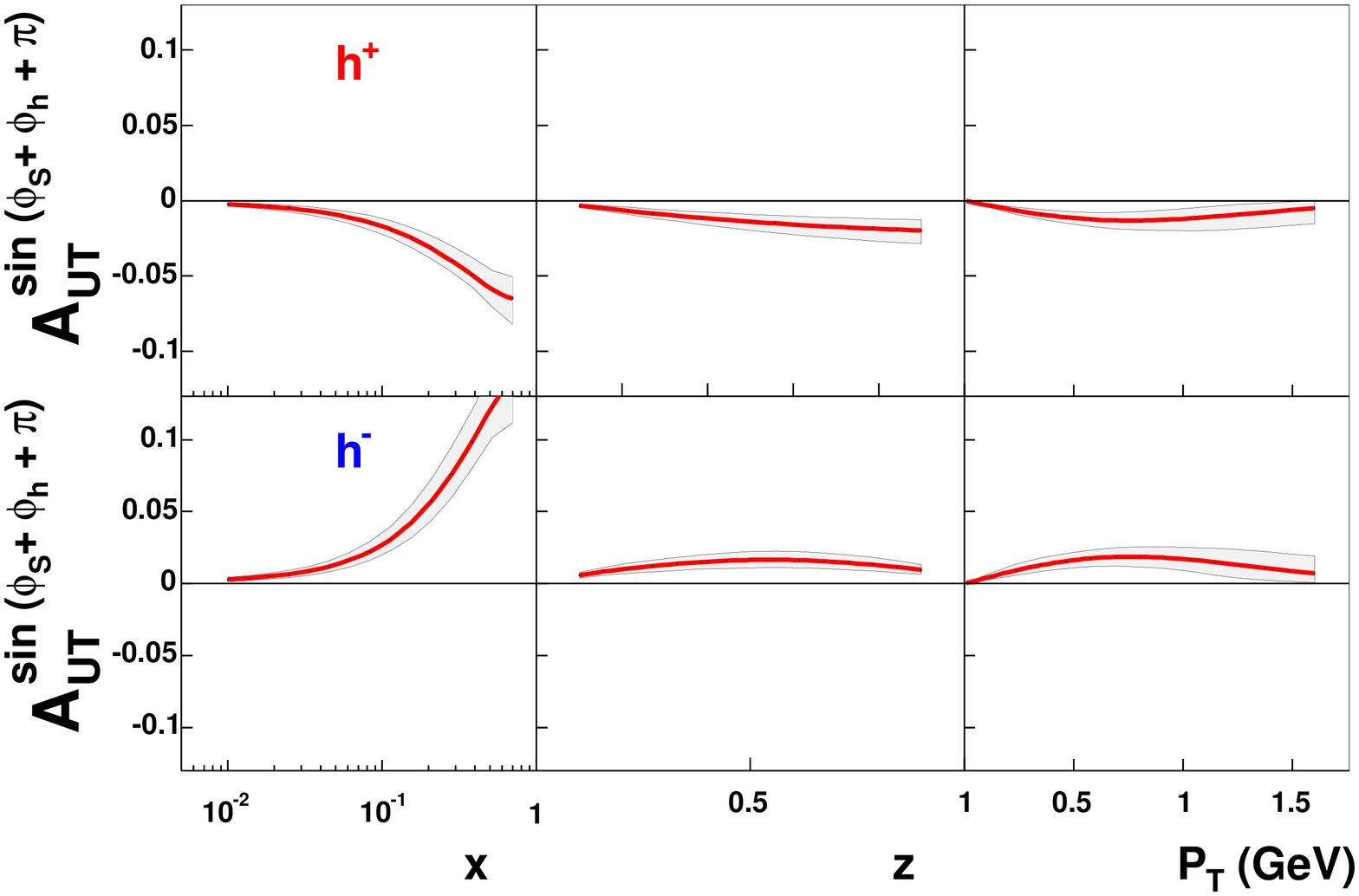} \hskip 2.5cm
\includegraphics[width=0.4\textwidth,bb= 10 140 540 660,angle=-90]
{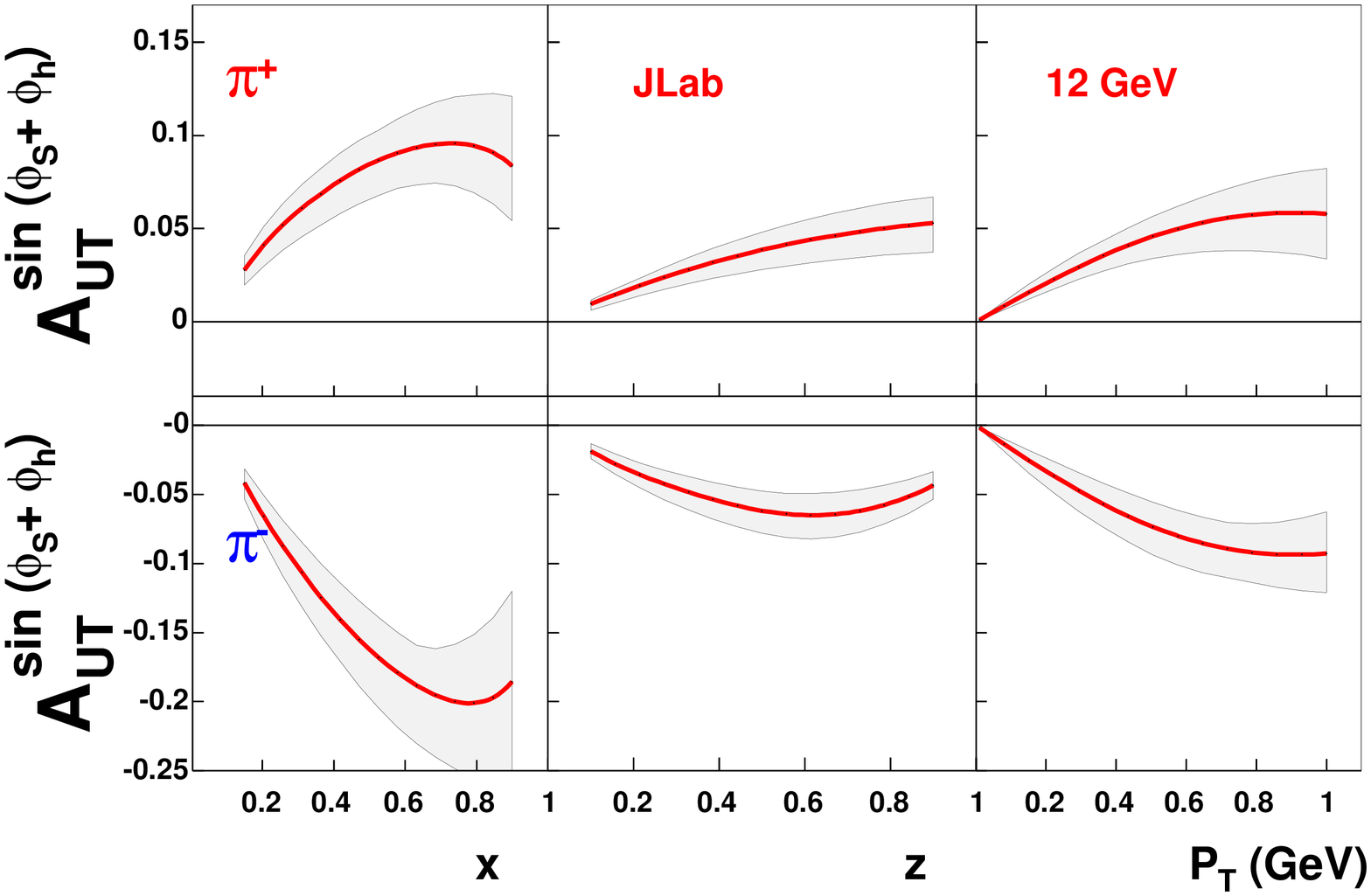}
\end{center}
\vskip -0.5cm
\caption{\label{fig:compass-proton}
Estimates of the single spin asymmetry $A_{_{UT}}^{\sin(\phi_h+\phi_S)}$
for COMPASS (left panel) and JLab (right panel) operating with proton target.}
\end{figure}

Finally, in Fig.~\ref{fig:compass-proton}, we show our estimates for COMPASS 
and JLab experiments operating on proton target. Notice that JLab results will 
give important information on the range of large $x-$values, which is left 
basically unconstrained by the present SIDIS data from HERMES and COMPASS. Our 
large$-x$ estimates are based on an extrapolation of the transversity 
distribution function into an unexplored region of $x$, and consequently must 
be taken with some care.  

The first $x$-moments of the transversity distribution -- related to the 
tensor charge, and defined as 
$\Delta_T q \equiv \int_0^1{\rm d} x \Delta_T q(x)$ 
-- are found to be $\Delta_T u = 0.59^{+0.14}_{-0.13}$, 
$\Delta_T d =-0.20^{+0.05}_{-0.07}$ at $Q^2 = 0.8$  GeV$^2$.

{From} this preliminary analysis, we conclude that the inclusion of the new 
data sets~\cite{Diefenthaler:2007rj,Alekseev:2008dn,Seidl:2008xc} 
improves our determination of the transversity distribution and Collins FFs, 
as can be seen by the reduced size of the present uncertainty bands 
(Fig.~\ref{fig:transv-coll}),  
and enables us to give more precise predictions for forthcoming 
experiments (see Fig.~\ref{fig:compass-proton}). The new distributions 
are compatible with the extraction of Ref.~\cite{Anselmino:2007fs} and 
close to some model predictions for the transversity distribution 
(see for example Ref.~\cite{Cloet:2007em}).  

\begin{footnotesize}

\end{footnotesize}

\end{document}